\documentclass[traditabstract,letter,a4paper]{aa} 
\usepackage{graphicx} 
\usepackage{txfonts}
\usepackage{natbib} 
\usepackage{amsfonts}
\usepackage{threeparttable}
\usepackage{booktabs}
\usepackage{wasysym}  

\begin{document}
   \title{Detection of CH$^+$ emission from the disc around \object{HD~100546}\thanks{Herschel is an ESA space observatory with
    science instruments provided by Principal Investigator
    consortia. It is open for proposals for observing time from the
    worldwide astronomical community.}}

   \subtitle{}

   \author{W.-F. Thi\inst{1}, F. M\'enard\inst{1}, G. Meeus\inst{2}, C. Martin-Za\"{i}di\inst{1}, P. Woitke\inst{3,4,5,1}, E. Tatulli\inst{1}, M. Benisty\inst{6}, I. Kamp\inst{7}, I. Pascucci\inst{8}, C. Pinte\inst{1}, C. A. Grady\inst{9}, S. Brittain\inst{10}, G.~J. White\inst{11,12} C. D. Howard\inst{13}, G. Sandell\inst{13}, C. Eiroa\inst{2}}

   \institute{ UJF-Grenoble 1 / CNRS-INSU, Institut de Plan\'{e}tologie et d'Astrophysique (IPAG) UMR 5274, Grenoble, F-38041, France\\
     \email{Wing-Fai.Thi@obs.ujf-grenoble.fr} \and Dep.\ de F\'isica
     Te\'orica, Fac.\ de Ciencias, UAM Campus Cantoblanco, 28049
     Madrid, Spain \and University of Vienna, Dept.\ of Astronomy,
     T{\"u}rkenschanzstr.~17, A-1180 Vienna, Austria \and School of
     Physics \& Astronomy, University of St.~Andrews, North Haugh,
     St.~Andrews KY16 9SS, UK \and UK Astronomy Technology Centre,
     Royal Observatory, Edinburgh, Blackford Hill, Edinburgh EH9 3HJ,
     UK \and Max Planck Institute for Astronomy, K\"{o}nigstuhl 17
     D-69117, Heidelberg, Germnay
\and Kapteyn
     Astronomical Institute, P.O. Box 800, 9700 AV Groningen, The
     Netherlands \and Space Telescope Science Institute, 3700 San
     Martin Drive, Baltimore, MD 21218, USA \and Eureka Scientific and
     Exoplanets and Stellar Astrophysics Lab, NASA Goddard Space
     Flight Center, Code 667, Greenbelt, MD, 20771, USA \and Clemson
     University, Clemson, SC, USA \and Department of Physics \&
     Astronomy, The Open University, Milton Keynes MK7 6AA, UK \and The
     Rutherford Appleton Laboratory, Chilton, Didcot, OX11 OQL, UK
     \and SOFIA-USRA, NASA Ames Research Center, Mailstop 211-3
     Moffett Field CA 94035 USA }

   \date{Received 2011; accepted 2011}

 
   \abstract {Despite its importance in the thermal balance of the gas
     and in the determination of primeval planetary atmospheres, the
     chemistry in protoplanetary discs remains poorly constrained with
     only a handful of detected species.  We observed the emission
     from the disc around the Herbig~Be star \object{HD~100546} with
     the {\sc PACS} instrument in the spectroscopic mode on board the
     {\sc Herschel Space Telescope} as part of the GaS in
     Protoplanetary Systems ({\sc GASPS}) programme and used archival
     data from the {\sc DIGIT} programme to search for the rotational
     emission of CH$^+$. We detected in both datasets an emission line
     centred at 72.16~$\mu$m that most likely corresponds to the
     $J$=5-4 rotational emission of CH$^+$. The $J$=3-2 and 6-5
     transitions are also detected albeit with lower confidence. Other
     CH$^+$ rotational lines in the {\sc PACS} observations are
     blended with water lines. A rotational diagram analysis shows
     that the CH$^+$ gas is warm at 323$^{+2320}_{-151}$~K with a mass
     of $\sim$3 $\times$ 10$^{-14}$-5 $\times$ 10$^{-12}$
     M$_\odot$. We modelled the CH$^+$ chemistry with the
     chemo-physical code {\sc ProDiMo} using a disc density structure
     and grain parameters that match continuum observations and near-
     and mid-infrared interferometric data. The model suggests that
     CH$^+$ is most abundant at the location of the disc rim at
     10-13~AU from the star where the gas is warm, which is consistent
     with previous observations of hot CO gas emission.  }

   {}
\authorrunning{Thi et al.}
\titlerunning{Detection of CH$^+$ in \object{HD~100546}}            

   \maketitle
%

\section{Introduction}

Planets are formed in discs around young stars. The building of
planets is intimately linked to the grain-growth process and the
disc's chemical evolution \citep{Bergin2007prpl.conf..751B}. The
chemical and temperature profiles of discs determine the chemical
composition of the gas that will eventually be incorporated into a
planet's atmosphere, once a solid $\sim$~10 Earth mass core is formed
\citep{Armitage2010apf}. Millimetre observations of molecules in discs
probe the outer cold part
(\citealt[e.g.]{Dutrey1997,vanZadelhoff2001,Thi2004A&A...425..955T,Semenov2005,Henning2010ApJ...714.1511H,Oberg2010ApJ...720..480O}). Minor
species located in the inner disc, such as C$_2$H$_2$ and HCN, were
observed with the {\sc Spitzer Space Telescope}
\citep{Carr2008Sci...319.1504C,Pascucci2009ApJ...702..724P,Lahuis2006ApJ...636L.145L,Pontoppidan2010ApJ...720..887P}. Gas-phase
carbon is believed to be locked into C$^+$, C, and CO. However, a
significant fraction of the carbon can be locked into CH and CH$^+$,
which are observed in the optical in diffuse clouds.\\  
\indent The {\sc Herschel Space Telescope} permits observations of
hydrides and their ions (OH, OH$^+$, CH, CH$^+$, ...), and warm light
molecular gas (e.g., H$_2$O), which all emit in the
far-infrared. Rotational lines of CH$^+$ at 300-500~K in the far-IR
have been observed by {\sc ISO} towards the \object{NGC7027} PDR
\citep{Cernicharo1997ApJ...483L..65C}. The isotopologue $^{13}$CH$^+$
$J$=1-0 was detected in absorption in the diffuse medium by
\citet{Falgarone2005ApJ...634L.149F}. The lowest rotational transition
was detected in emission in the Orion Bar
\citep{Naylor2010A&A...518L.117N,Habart2010A&A...518L.116H} and in
absorption towards \object{DR 21} \citep{Falgarone2010A&A...518L.118F}
with {\sc Herschel}. The neutral counterpart CH has also been detected
in the diffuse medium by {\sc Herschel}
\citep{Qin2010A&A...521L..14Q,Gerin2010A&A...521L..16G}. In this
letter, we discuss the detection of emission lines with Herschel that
we assign to the rotational transition of CH$^+$ from the disc around
the 10 Myrs old Herbig~Be star \object{HD~100546}
($M_*$=~2.2~M$_\odot$, $L_*$=~27~L$_\odot$, $T_{\mathrm
  {eff}}$=~10500~K) located at 103~pc
\citep{vandenancker1997A&A...324L..33V}. \object{HD~100546} shows a
very rich gas and a solid spectrum in the infrared and millimetre
ranges
\citep{Brittain2009ApJ...702...85B,sturm2010A&A...518L.129S,Panic2010A&A...519A.110P,Malfait1998A&A...332L..25M}.
\begin{figure*}
\centering
\resizebox{\hsize}{!}{\includegraphics[angle=90]{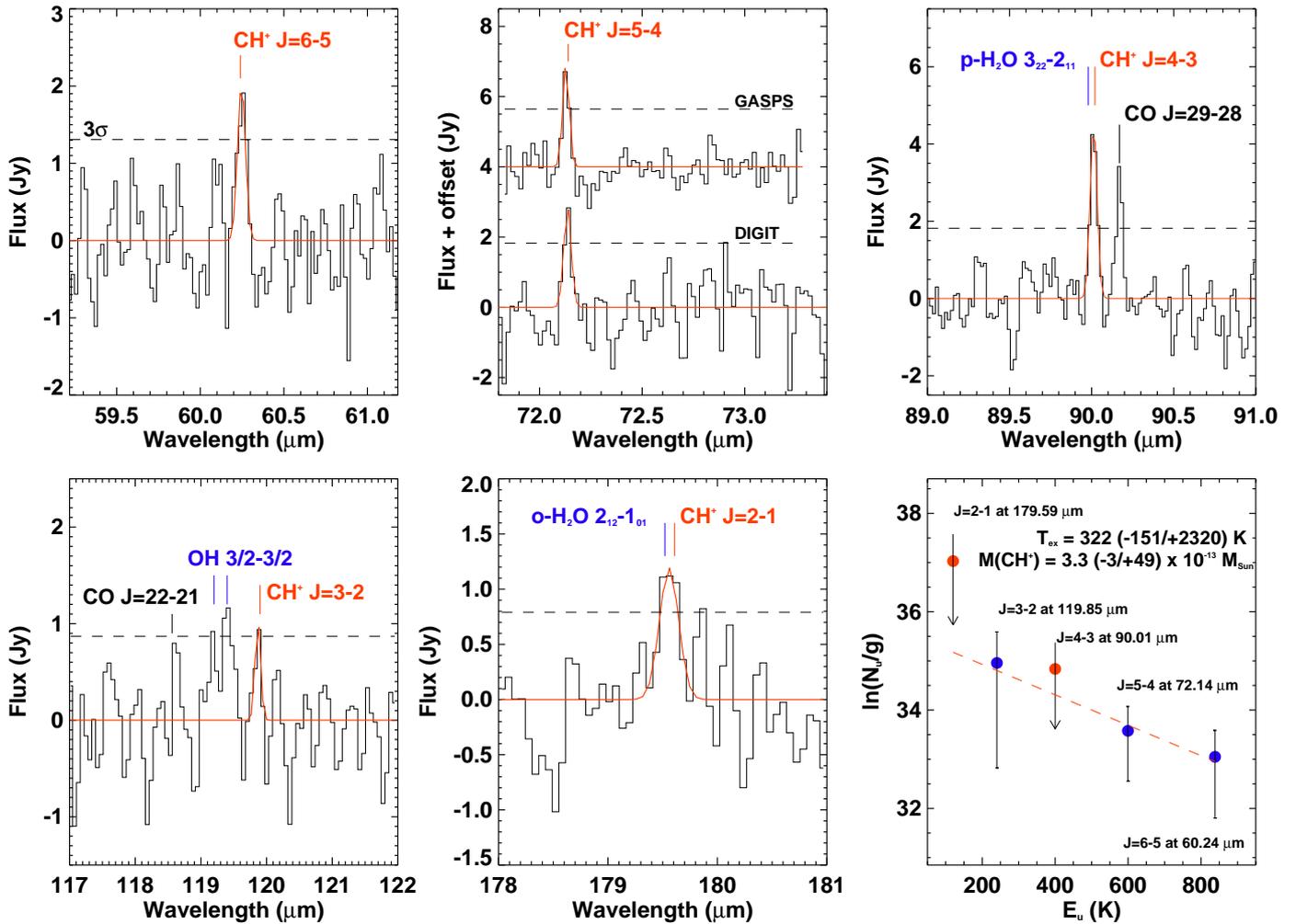}}
\caption{Continuum-subtracted {\sc Herschel-PACS} spectra of
  \object{HD~100546} around the CH$^+$ $J$=6--5, 4--3, 3--2, 2--1
  (DIGIT data), and $J$=5-4 (GASPS and DIGIT data) lines. The
  3$\sigma$ statistical error levels do not include the 30\%
  calibration uncertainty. A rotational diagram using those
  transitions is plotted in the lower-right panel. The errors in the
  diagram are the quadratic sum of all the statistical and calibration
  errors ($err_{tot}=\sqrt{(3\sigma_{stat})^2+(0.3F_{obs})^2}$). Blended lines are considered as upper limits (red filled
  dots).}
  \label{fig_chplus_hd100546_Herschel}  
\end{figure*}  
   
\section{Herschel observations and rotational diagram}\label{herschel_observation}
We observed \object{HD~100546} with {\sc PACS} \citep{Poglitsch2010}
on board the {\sc Herschel Space Telescope} \citep{Pilbratt2010} in
the 72-73~$\mu$m spectroscopic scan range ($\Delta v$=~163~km
s$^{-1}$), as part of the open-time key programme {\sc GASPS} (GaS in
Protoplanetary Systems, obsid 1342188437), which aims at a systematic
study of gas and dust in discs around stars aged between 0.5 and 30
Myrs.  The observations were reduced using the standard pipeline in
{\sc HIPE 4.2.} We detected a line at ~72.14 $\mu$m that we assign to
the rotational $J$=5-4 emission of CH$^+$ (see Fig.\
\ref{fig_chplus_hd100546_Herschel}). We also detected the same line in
the science demonstration data spectrum taken by the DIGIT team
\citep{sturm2010A&A...518L.129S}, together with the $J$=6-5
(60.24~$\mu$m) and $J$=3-2 (119.68~$\mu$m) lines but with a lower
signal-to-noise ratio.  It was not possible to strengthen our analysis
with detection of either the lowest rotational transition $J$=1-0,
which lies in the {\sc SPIRE/HIFI} range, or the $J$=2-1 and $J$=4-3
lines in the {\sc DIGIT} dataset, which are blended with water
emission lines at the {\sc PACS} resolution. The ortho H$_2$O
2$_{12}$-1$_{01}$ emission at 179.52~$\mu$m is blended with the CH$^+$
$J$=2-1 line at 179.61~$\mu$m, and the para-H$_2$O 3$_{22}$-2$_{11}$
at 89.98~$\mu$m is blended with the CH$^+$ $J$=4-3 line at
90.02~$\mu$m. The flux calibration accuracy of the {\sc Herschel-PACS}
instrument is currently 30\% in addition to the statistical noise. In
addition, flux uncertainties are introduced by difficulties in
determining the continuum level in the {\sc DIGIT} spectra. A new
reduction with HIPE 7.0.822, which includes division by the spectral
response function and flatfielding, does not improve the noise level
in the continuum.  \indent The line ratio $\mathrm{[O~{\sc I}]}$
63~$\mu$m/CH$^+$ $J$=5-4 is 80. We fitted a Gaussian profile to each
line. The observed CH$^+$, $\mathrm{[O~{\sc I}]}$, $\mathrm{[C~{\sc
    II}]}$ \citep{sturm2010A&A...518L.129S}, and CO 3--2
\citep{Panic2010A&A...519A.110P} fluxes are all reported in
Table~\ref{tab_chplus_fluxes}.\\
\indent We constructed a rotational diagram by assuming LTE population
and optically thin emission (see lower right panel of
Fig.~\ref{fig_chplus_hd100546_Herschel}). The level energies, line
frequencies, and Einstein spontaneous emission coefficients were
computed by \citet{Muller2010A&A...514L...6M} based on the
spectroscopic analysis of \citet{Amano2010ApJ...716L...1A} . We
analysed the diagram using censored data regression and bootstrapping
methods.  We derived an excitation temperature of
323$^{+2320}_{-150}$~K, a disc-averaged CH$^+$ column density of
4.3$^{+63.8}_{-3.9}$~$\times$~10$^{12}$~cm$^{-2}$, consistent with the
upper limit found in observing CH$^+$ at 4232.7~\AA\ (Martin-Za\"{i}di
et al., submitted), and a lower limit to the CH$^+$ mass of
3.3$^{+48.9}_{-3.0}$~$\times$~10$^{-13}$ M$_\odot$. The CH$^+$
emitting gas is warm, which suggests that CH$^+$ is located within
100~AU of the star. The CH$^+$ abundance is
$\ge$~3.9~$\times$~10$^{-9}$ relative to an upper limit of warm
molecular gas from the VISIR H$_2$ $S$(1) 17~$\mu$m observation of
M($T_{\mathrm{gas}}$=300~K)$\leq$~8.4~$\times$~10$^{-5}$ M$_\odot$
\citep{Martin2010A&A...516A.110M}.
\begin{center}      
\begin{table}
  \caption{Observed and modelled line fluxes}\label{tab_chplus_fluxes}
\begin{tabular}{lrrrr}
  \toprule
  Transition  &\multicolumn{1}{c}{$\lambda$}&\multicolumn{1}{c}{obs.\tnote{1}} & \multicolumn{1}{c}{model 1} & \multicolumn{1}{c}{model 2} \\
  &\multicolumn{1}{c}{($\mu$m)}&\multicolumn{3}{c}{(10$^{-17}$ W m$^{-2}$)}\\
  \noalign{\smallskip}   
  $\mathrm{[O~{\sc I}]}$  $^3$P$_1-^3$P$_2$ & 63.19     &  554.37~$\pm$~167$^a$ &  785 & 763\\ 
  $\mathrm{[O~{\sc I}]}$  $^3$P$_0-^3$P$_1$ &  145.54     & 35.70~$\pm$~11.4 &  35.7  & 42.7\\ 
  $\mathrm{[C~{\sc II}]}$ $^2$P$_{3/2}-^2$P$_{1/2}$ & 157.75  &  31.87~$\pm$~10.0  & 12.7 & 17.2\\ 
  p-H$_2$O $3_{22}-2_{11}$ & 89.90 & $\leq$~14.32~$\pm$~5.6$^{b}$ & 4.6  & 7.4  \\
  CH$^+$       $J=4-3$ & 90.02  & $\leq$~14.32~$\pm$~5.6$^{b}$ & 3.7  & 6.3  \\
  CH$^+$       $J=6-5$ & 60.25  &  10.32~$\pm$~5.7 & 4.2 & 8.9 \\ 
  CH$^+$       $J=5-4^c$ &  72.14  & 6.86~$\pm$~3.4 & 3.3 & 6.7 \\ 
  CH$^+$       $J=3-2$ &   119.87 &   2.16~$\pm$~1.1 & 1.8 & 2.9  \\ 
  CO            $J=3-2^d$&   866.96   & 0.10~$\pm$~0.03       & 0.08 & 0.24\\
  \bottomrule
\end{tabular}
$^a$ The total (3$\sigma$ statistical + 30\% calibration) errors are given for the {\sc PACS}
observations. $^b$ The blended H$_2$O+CH$^+$ line is detected. $^c$ The {\sc PACS} values are from the {\sc DIGIT} programme \citep{sturm2010A&A...518L.129S} except for the CH$^+$ $J=5-4$ flux ({\sc GASPS} programme). $^d$ Data from \citet{Panic2010A&A...519A.110P} with 3$\sigma$ error.
\end{table}
\end{center} 
\begin{center}
\begin{table}
  \caption{Model parameters.}\label{tab_chplus_parameters}
\begin{tabular}{lrrr}
  \toprule
  \multicolumn{4}{c}{{\sc MCFOST$^a$}}\\
  &     inner disc      & surf. layer & outer disc\\
  \noalign{\smallskip}   
  Inner radius $R_{\mathrm{in}}$ (AU)   & 0.24   &   13  & 13\\
  Outer radius $R_{\mathrm{out}}$ (AU)   &  4     &   50  & 500\\
  Surf.\ density exponent $q$ & 1 & 0.5 & 1.125\\
  Scale height $H_{\mathrm{100AU}}$ (AU)& 6 & 14$^a$ & 14$^a$ \\
  Scale height exponent $\beta$ & 1.0 & 1.0 & 1.0\\
  Total dust mass $M_{\mathrm{dust}}$  (M$_\odot$) & 1.75(-10)$^b$ & 3(-7) & 4.3(-4)\\
  Dust mass  ($a \leq $ 1mm, M$_\odot$) & 1.75(-10) & 3(-7) & 1.3(-4) \\
  Min.\ grain radius $a_{\mathrm{min}}$ ($\mu$m) & 0.1 & 0.05 & 1\\
  Max.\ grain radius $a_{\mathrm{max}}$ ($\mu$m) & 5 & 1 & 10$^4$\\
  Grain power law index $p$ & 3.5 & 3.5 & 3.5\\
  Silicate grain density  (g cm$^{-3}$) & 3.0 & 3.0 & 3.0\\
  \noalign{\smallskip}   
  \hline
  \noalign{\smallskip}   
  \multicolumn{4}{c}{{\sc ProDiMo$^c$}}\\
  ISM UV field      ($\chi$, Habing)  & \multicolumn{3}{c}{1.0} \\
  viscosity         ($\alpha$)        & \multicolumn{3}{c}{0.0}  \\
  Non-thermal speed   (km s$^{-1}$) & \multicolumn{3}{c}{0.15} \\
  Disc inclination  (\degr)             & \multicolumn{3}{c}{42} \\
  UV excess            & \multicolumn{3}{c}{0.013} \\
  UV power-law index    & \multicolumn{3}{c}{6.5} \\
  Cosmic ray flux    $\zeta$(s$^{-1}$)         & \multicolumn{3}{c}{1(-17)}\\
  PAH C$_{150}$H$_{30}$ mass          (M$_\odot$)  & \multicolumn{3}{c}{1.8(-7)} \\
  Gas mass (M$_\odot$) & \multicolumn{3}{c}{5(-4) (model 1)}\\
  Gas mass (M$_\odot$) & \multicolumn{3}{c}{1(-3) (model 2)} \\
  \bottomrule
\end{tabular}
$^a$ Values taken from \citet{Benisty2010A&A...511A..75B} and Tatulli et al.\ (2011) except for the scale height $H_{\mathrm{100AU}}$ at 100~AU. $^b$ $\alpha(-\beta)$ means $\alpha$~$\times$~10$^{-\beta}$. $^c$ This work.
\end{table}
\end{center}
\vspace{-2cm}
\section{Modelling the physics and chemistry of CH$^+$}\label{prodimo_model}
We first fitted the spectral energy distribution (SED) and infrared
interferometric {\sc VLT-AMBER} and {\sc MIDI} data to constrain the
dust properties and the disc structure with a hydrostatic disc using
the {\sc MCFOST} radiative transfer code
\citep{Pinte2006A&A...459..797P,Pinte2009A&A...498..967P}. The details
of the fit are discussed in \citet{Benisty2010A&A...511A..75B} and
Tatulli et al.\ (2011, submitted). The disc is composed of three
parts: the inner disc, the outer disc, and an upper layer on the top
of the outer disc (see Table~\ref{tab_chplus_parameters},
Fig.~\ref{fig_chplus_hd100546_density_profile}, and the sketch in
\citealt{Benisty2010A&A...511A..75B}). The inner and outer discs are
separated by a gap
\citep{Bouwman2003A&A...401..577B,Grady2005ApJ...620..470G}. The fit
to the SED in Tatulli et al.\ (2011) suggests a scale height value of
10 AU at 100~AU, which is consistent with the scattered-light images,
but equally good fits can also be obtained with other values.
Therefore we tried several disc structures by varying the scale height
$H_{\mathrm{100AU}}$. Using the disc structures derived from the fit
to the SED, we modelled the gas chemistry and radiative transfer in the
\object{HD~100546} disc with the {\sc ProDiMo} code
\citep{Woitke2009A&A...501..383W,Kamp2010A&A...510A..18K}. The {\sc
  ProDiMo} code computes the chemistry and heating-cooling balance of
the gas self-consistently. We assumed a non-viscous gas ($\alpha$=0)
and a PAH mass (C$_{150}$H$_{30}$) that is consistent with the strong
PAH emissions in the IR \citep{Keller2008ApJ...684..411K}. Large PAHs
($\geq$~100 carbon atoms) exhibit a strong feature between 10.9 and
11.3~$\mu$m \citep{Bauschlicher2009ApJ...697..311B} as seen in the
spectrum of \object{HD~100546}. The amount of PAH influences both the
thermal state of the disc and the ionisation state of the upper
layers.  The non-thermal line width is constrained by the resolved CO
3-2 profile. We used a dereddened FUSE and IUE UV spectrum as input to
the gas modelling \citep{Martin2008A&A...484..225M}. We also ran a
{\sc ProDiMo} model where the disc vertical structure is calculated
according to the gas
pressure and found that the height of the rim at 13~AU is consistent with the fixed three-part structure.\\
The main CH$^+$ formation reaction C$^+$ + H$_2$ $\rightarrow$ CH$^+$
+ H (reaction 1) is endothermic by 0.398 eV (4537K).  The reaction is
efficient only at a few 100 K found in shocked or turbulent regions or
when H$_2$ is vibrationally excited
\citep{Hierl1997JChPh.10610145H,Agundez2010ApJ...713..662A}. The rate
for reaction 1 in \citet{Herbst1981ApJ...245..529H} underestimates the
experimental data by a factor of 2-3, and the {\sc UMIST 2006} rate
\citep{Wooddall2007A&A...466.1197W} is a factor 7--8 lower than the
experimental values.  Initial models with a low rate for reaction 1
underpredict the CH$^+$ abundance by more than a factor 10. We adopted
the value suggested by \citet{Hierl1997JChPh.10610145H}: $k_1(T)=(7.4
\pm 0.8)\times 10^{-10} e^{-4537/T} \mathrm{cm}^{3}\ \mathrm{s}^{-1}$.
We also take the enhanced rate for vibrationally-excited molecular
hydrogen into account by assuming that the rate is equal to the
Langevin collision value of 1.6 $\times$10$^{-9}$ cm$^{3}$ s$^{-1}$
\citep{Agundez2010ApJ...713..662A}. The alternative radiative
association reaction $\mathrm{C}^+ + \mathrm{H} \rightarrow
\mathrm{CH^+} +{\mathrm{h}}\nu$ has an extremely low rate of
$\sim$~10$^{-16}$-10$^{-17}$ cm$^{-3}$ s$^{-1}$
\citep{Barinovs2006ApJ...636..923B}. The warm and UV-excited gas at
the disc atmosphere around a Herbig~AeBe star renders reaction 1
fast. CH$^+$ reacts with H$_2$ to form CH$_2^+$, which in turn reacts
with H$_2$ to form CH$^+_3$. As both exothermic reactions are fast, an
enhanced abundance of CH$^+$ is accompanied by a high abundance of
CH$^+_2$ and CH$^+_3$, which have not been detected yet. CH$^+$ can
also dissociatively recombine with an electron into C$^+$ and H or
exchange its charge with PAHs. CH$^+$ is also destroyed upon
absorption of UV photons. The absorption cross-section peaks at
$\sim$950 \AA.\ Therefore, the photodissociation around a B9.5V star
without strong UV excess due to accretion is not as fast as with an
interstellar UV field. The photodissociation and photoionisation rates
are computed inside the disc using cross-sections taken from
\citet{vanDishoeck2006FaDi..133..231V}. The steady-state chemistry
includes 188 gas and solid species.\\
\indent We ran several models with disc gas mass between 10$^{-4}$ and
10$^{-2}$ M$_\odot$ and H$_{\mathrm{100AU}}$ between 10 and 20~AU
(Fig.~\ref{fig_chplus_hd100546_dust_temperature}). In
Table~\ref{tab_chplus_fluxes} we show the three detected line fluxes
for CH$^+$ out of the 15 rotational transitions in the model, the OI
and CII fine-structure lines, and the CO 3--2 transition for our
preferred models, which have a disc mass of (0.5--1)~$\times$~10$^{-3}$
M$_\odot$ and H$_{\mathrm{100AU}}=14$~AU, for the outer disc. All the
modelled lines agree within a factor 2 with the observations.  In
particular, the sum of the p-H$_2$O $3_{22}-2_{11}$ and CH$^+$ $J=4-3$
fluxes is consistent with the observed value. The collision rates of
CH$^+$ with electrons were computed by
\citet{Lim1999MNRAS.306..473L}. In the absence of published collision
rates with H$_2$, we scaled the collision rates with He by 1.39 for
$J\leq6$ transitions
\citep{Hammami2009A&A...507.1083H,turpin2010A&A...511A..28T}.

\begin{figure}
\centering
\resizebox{\hsize}{!}{\includegraphics[angle=0]{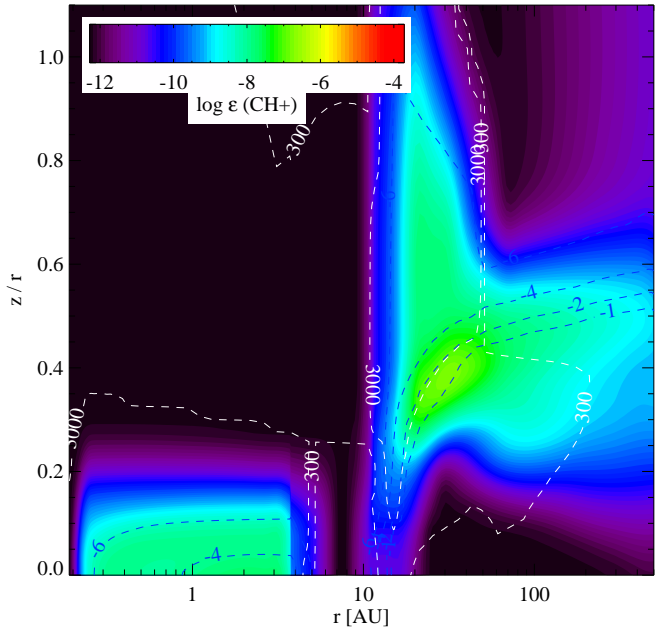}}
\caption{CH$^+$ abundance structure in the \object{HD~100546} disc
  with respect to the number of H-nuclei. The H$_2$ abundance relative
  to total H-nuclei abundance contours in log-scale are overplotted in
  blue, while the $T_{\mathrm{gas}}$=300 and 3000K contours are
  overplotted in white. The line-of-sight CH$^+$ observations in the
  UV would cross the disc at z/r=0.9 (42 \degr).}
  \label{fig_chplus_hd100546_density_abundance}  
\end{figure} 
\section{Discussion}\label{discussion}

The models suggest that CH$^+$ is mostly located on the rim (10-13~AU)
of the outer disc and at the disc surface
(Fig.~\ref{fig_chplus_hd100546_density_abundance}). Although CH$^+$ is
abundant in the tenuous inner disc, its amount is insufficient to
contribute to the 72.14~$\mu$m emission. CH$^+$ is located where the
gas is mostly heated by collisional de-excitations of nascent and
UV-pumped H$_2^*$. Part of the CH$^+$ is formed by the reaction with
excited H$_2^*$, but the main fraction comes from the reaction of
C$^+$ with thermal H$_2$ at $T_{\mathrm{gas}}>$~400~K. CH$^+$ reaches
a maximum abundance of $\sim$~10$^{-7}$ just behind the rim and has a
total mass of 10$^{-13}$-10$^{-12}$ M$_\odot$. The disc average CH$^+$
abundance is (0.7-7)~$\times$~10$^{-9}$ relative to H-nuclei,
consistent with the simple analysis in
Sect.~\ref{herschel_observation}. The high CH$^+$ abundance in the rim
region and upper disc layers where hydrogen is in atomic form does not
affect the C$^+$ abundance much, which reaches 10$^{-4}$.  However
below the disc atmosphere, H$_2$ self-shielding is efficient, and
C$^+$ is converted rapidly into CH$^+$. The formation reaction of
CH$^+$ (C$^+$+H$_2$) competes with the recombination reaction of C$^+$
to form neutral carbon.\\
\indent Warm OH \citep{sturm2010A&A...518L.129S} and fluorescent
ro-vibrational CO emission
\citep{Brittain2009ApJ...702...85B,vanderPlas2009A&A...500.1137V} have
also been detected in \object{HD~100546} . OH is mostly formed by the
reaction of atomic oxygen O with thermally hot H$_2$ and vibrationally
excited H$_2^*$. The presence of hot CO, CH$^+$ and OH supports the
idea that hot and excited gas chemical reactions occur on disc
surfaces and at the inner rim.  \vspace{-0.4cm}
\section{Conclusions}\label{conclusions}
We detected the $J$=5-4 rotational transition of CH$^+$ at
72.16~$\mu$m in two {\sc Herschel} datasets.  We also tentatively
detected other lines from CH$^+$ at 60.25 and 119.87 $\mu$m.  Other
CH$^+$ lines are blended with water lines. We modelled the CH$^+$ line
fluxes using the most recent chemical and collisional rates. Searches
for CH$^+$ in other Herbig~AeBe discs are warranted to test whether
the presence of CH$^+$ is unique to \object{HD~100546} or if it is
related to the presence of a high temperature rim of gas, or both.
\vspace{-0.1cm}
\begin{acknowledgements}
  The Grenoble group acknowledges PNPS, CNES, and ANR (contract
  ANR-07-BLAN-0221).  C. Eiroa and G. Meeus are partly supported by
  the Spanish grant AYA 2008-01727. C. Pinte acknowledges the funding
  from the EC 7$^{th}$ Framework Programme (PIEF-GA-2008-220891,
  PERG06-GA-2009-256513).  I. Pascucci, M. Grady, S. Brittain,
  C. Howard, and G. Sandell acknowledge NASA/JPL.
\end{acknowledgements}
\vspace{-0.5cm}
\bibliographystyle{aa}
\bibliography{protoplanetary_disks}
\appendix
\section{Disc density and gas temperature structure}
\begin{figure}
\centering
\resizebox{\hsize}{!}{\includegraphics[angle=0]{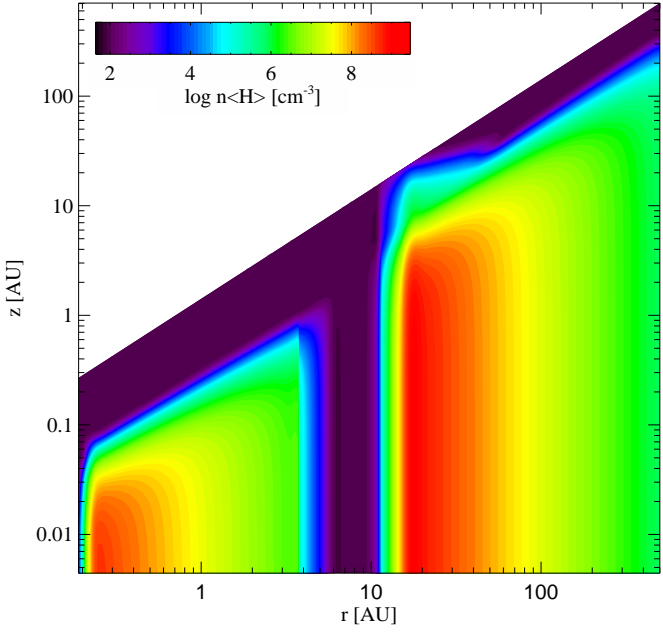}}
\caption{The input disc density structure for model 2 constrained by
  the fit to the SED (see Tatulli et al.\ 2011 for details). The gap
  is filled with a gas at density of $n_{\mathrm{H}}$=100 cm$^{-3}$.}
  \label{fig_chplus_hd100546_density_profile}  
\end{figure}

\begin{figure}
\centering
\resizebox{\hsize}{!}{\includegraphics[angle=0]{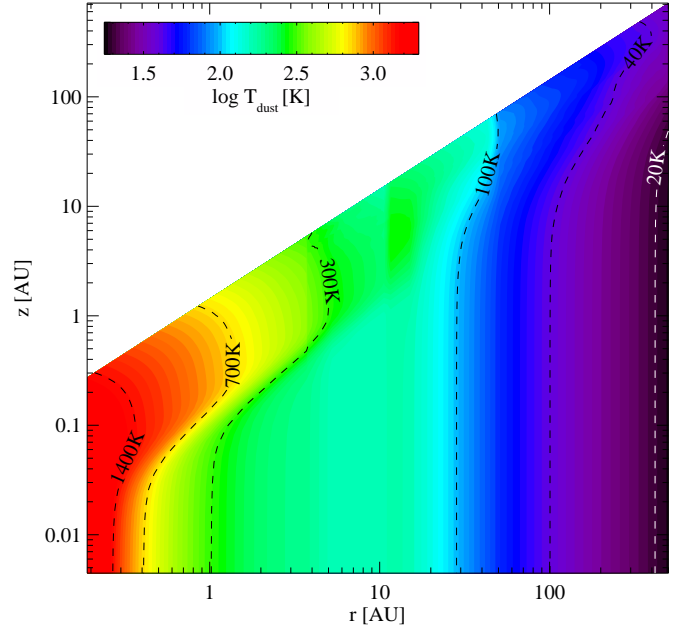}}
\caption{The disc dust temperature profile computed by {\sc MCFOST} for model 2.}
  \label{fig_chplus_hd100546_dust_temperature}  
\end{figure}

\begin{figure}
\centering
\resizebox{\hsize}{!}{\includegraphics[angle=0]{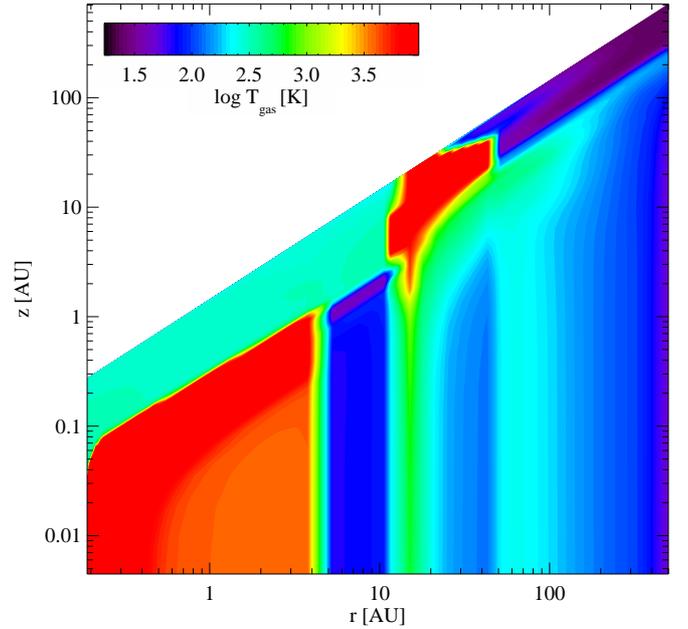}}
\caption{The disc gas temperature profile computed by {\sc ProDiMo} for model 2.}
  \label{fig_chplus_hd100546_gas_temperature}  
\end{figure}

We provide for completeness the disc density and the dust and gas
temperature structures for model 2
(Fig.~\ref{fig_chplus_hd100546_density_profile},
~\ref{fig_chplus_hd100546_dust_temperature}, and
~\ref{fig_chplus_hd100546_gas_temperature}).

\end{document}